\documentclass[aps,prl,twocolumn,groupedaddress,showpacs]{revtex4}
\usepackage[dvips]{graphicx}
\newcommand{\BlackHat}{{\sc BlackHat}}
\newcommand{\SHERPA}{{\sc SHERPA}}
\newcommand{\AMEGIC}{{\sc AMEGIC++}}

\newif\ifdraft
\drafttrue
\newif\ifpreprint
\preprinttrue

\def\fig#1{fig.~{\ref{#1}}}

\def\tab#1{table~{\ref{#1}}}

\def\e{\epsilon}
\def\bom#1{{\mbox{\boldmath $#1$}}}
\def\Wj{$W\,\!+\,1$}
\def\Wjj{$W\,\!+\,2$}
\def\Wjjj{$W\,\!+\,3$}
\def\Wjx{$W\,\!+\,1,2$}
\def\Wjjx{$W\,\!+\,2,3$}
\def\Wjjjx{$W\,\!+\,1,2,3$}
\def\Wjn{$W\,\!+\,n$}
\def\ETsl{{\s E}_T}

\newbox\charbox
\newbox\slabox
\def\s#1{{      
        \setbox\charbox=\hbox{$#1$}
        \setbox\slabox=\hbox{$/$}
        \dimen\charbox=\ht\slabox
        \advance\dimen\charbox by -\dp\slabox
        \advance\dimen\charbox by -\ht\charbox
        \advance\dimen\charbox by \dp\charbox
        \divide\dimen\charbox by 2
        \raise-\dimen\charbox\hbox to \wd\charbox{\hss/\hss}
        \llap{$#1$}
}}

\begin{document}

\ifpreprint
\hbox{
UCLA/09/TEP/35 $\null\hskip .7cm \null$
MIT-CTP~4013 $\hskip .7cm \null$
Saclay-IPhT-T09/019 $\hskip .7cm \null$
IPPP/09/08 $\hskip .7cm \null$
SLAC--PUB--13539
}
\fi

\title{Precise Predictions for $\bom{W}$\,+\,3 Jet Production 
at Hadron Colliders}

\author{C.~F.~Berger${}^{a}$, Z.~Bern${}^b$,
L.~J.~Dixon${}^c$, F.~Febres Cordero${}^b$, D.~Forde${}^{c}$,
T.~Gleisberg${}^c$,  H. Ita${}^b$, 
D.~A.~Kosower${}^d$ and D.~Ma\^{\i}tre${}^e$}%
\affiliation{\centerline{${}^a${Center for Theoretical
Physics, Massachusetts Institute of Technology,
      Cambridge, MA 02139, USA}} \\
\centerline{${}^b$Department of Physics and Astronomy, UCLA, Los Angeles, CA
90095-1547, USA} \\
\centerline{${}^c$SLAC National Accelerator Laboratory, Stanford University,
             Stanford, CA 94309, USA} \\
\centerline{${}^d$Institut de Physique Th\'eorique, CEA--Saclay,
          F--91191 Gif-sur-Yvette cedex, France}\\
\centerline{${}^e$Department of Physics, University of Durham,
          DH1 3LE, UK}
}

\begin{abstract}
We report on the first next-to-leading order QCD computation of
\Wjjj-jet production in hadronic collisions including all partonic
subprocesses.  We compare the results with CDF data from the Tevatron,
and find excellent agreement.  The renormalization and factorization
scale dependence is reduced substantially compared to leading-order
calculations.  The required one-loop matrix elements are computed
using on-shell methods, implemented in a numerical program, \BlackHat.
We use the \SHERPA{} package to generate the real-emission
contributions and to integrate the various contributions over phase
space.  We use a leading-color (large-$N_c$) approximation for the
virtual part, which we confirm in \Wjx-jet production to be valid to
within three percent.  The present calculation demonstrates the
utility of on-shell methods for computing next-to-leading-order
corrections to processes important to physics analyses at the Large
Hadron Collider.
\end{abstract}

\pacs{12.38.-t, 12.38.Bx, 13.87.-a, 14.70.-e, 14.70.Fm, 11.15.-q,
11.15.Bt, 11.55.-m \hspace{1cm}}

\maketitle


Particle physicists have long anticipated the discovery of new physics
beyond the Standard Model at the Large Hadron Collider (LHC) at CERN.
In many channels, discovering, understanding, and measuring new
physics signals will require quantitatively reliable predictions for
Standard Model background processes.  Next-to-leading order (NLO)
calculations in perturbative QCD are crucial to providing such
predictions.  Leading-order (LO) cross sections suffer from large
normalization uncertainties, up to a factor of two in complex
processes.  NLO corrections typically reduce the uncertainties to
10--20\%~\cite{LesHouches}.

The production of a vector boson in association with multiple jets of
hadrons is an important process. It forms a background to Standard
Model processes such as top quark production, as well as to searches
for supersymmetry.  Here we present the first NLO computation of
\Wjjj-jet production that can be compared directly to data, namely CDF
results~\cite{WCDF} from the Tevatron.

The development of methods for computing high-multiplicity processes
at NLO has involved a dedicated effort over many years, summarized in
ref.~\cite{LesHouches}.  The longstanding bottleneck to NLO
computations with four or more final-state objects---including
jets---has been in evaluating one-loop (virtual) corrections.
Feynman-diagram techniques suffer from a rapid growth in complexity as the
number of legs increases; in QCD, NLO corrections to processes with four final
state objects have been limited to the case of all external
quarks~\cite{BDDP}. On-shell
methods~\cite{UnitarityMethod,%
Zqqgg,BCFUnitarity,BCFW,Bootstrap,Genhel,OPP,Forde,OnShellReview}, in
contrast, do not use Feynman diagrams, but rely on the analyticity and
unitarity of scattering amplitudes to generate new amplitudes from
previously-computed ones.  Such methods scale extremely well as the
number of external legs increases~\cite{Genhel,BlackHatI,GZ}, offering
a solution to these difficulties.

In an on-shell approach, terms in a one-loop amplitude containing
branch cuts are computed by matching the unitarity cuts (products of
tree amplitudes) with an expansion of the amplitude in terms of a
basis of scalar integrals~\cite{UnitarityMethod}.  Recent
refinements~\cite{BCFUnitarity,OPP,Forde,Badger}, exploiting
complexified loop momenta, greatly enhance the effectiveness of
generalized (multiple) cuts~\cite{Zqqgg}.  Evaluating the 
cuts in four dimensions allows the use of compact
forms for the tree amplitudes which enter as ingredients.
This procedure drops rational terms, which could be computed by
evaluating the cuts in $D$ dimensions~\cite{DdimUnitarity}. 
One may also obtain the rational terms using
on-shell recursion, developed by Britto, Cachazo, Feng and Witten
at tree level~\cite{BCFW}, and extended to loop level in
refs.~\cite{Bootstrap,Genhel}.  

Within the \BlackHat{} program~\cite{BlackHatI}, we
determine coefficients of scalar integrals using Forde's
analytic approach~\cite{Forde}, also incorporating elements from the
approach of Ossola, Papadopoulos and Pittau (OPP)~\cite{OPP}.  
For the rational terms, we have implemented both loop-level 
on-shell recursion and a massive continuation approach
(related to $D$-dimensional unitarity) along the lines
of Badger's method~\cite{Badger}.  The on-shell
recursion code is faster at present, so we use it here.  
The requisite speed and numerical stability of
\BlackHat{} have been validated for one-loop six-, seven- and
eight-gluon amplitudes~\cite{BlackHatI}, and for leading-color
amplitudes for a vector boson with up to five partons~\cite{ICHEPBH},
required for the present study. (A subsequent computation of
one-loop matrix elements needed for \Wjjj-jet production using
$D$-dimensional generalized unitarity within the OPP formalism was
described in ref.~\cite{W3EGKMZ}.)  Other numerical programs along
similar lines are presented in refs.~\cite{CutTools,GZ}.

\begin{figure}[t]
\includegraphics[clip,scale=0.60]{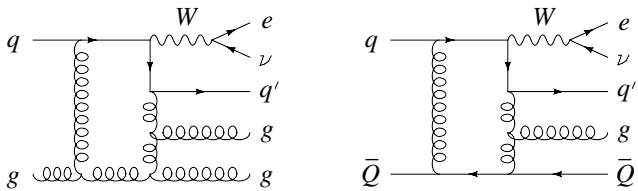}

\caption{Sample diagrams for the seven-point amplitudes
$q g \rightarrow e \nu \, q' \! g g$ and 
$q \bar Q \rightarrow e \nu \, q' \! g \bar Q$.  The $e \nu$
pair couples to the quarks via a $W$ boson.}
\label{lcdiagramsFigure}

\end{figure}

To speed up the evaluation of the virtual cross section, we make use
of a leading-color (large-$N_c$) approximation for the finite parts 
of the one-loop amplitudes, keeping the exact color dependence 
in all other parts of
the calculation.  Such approximations have long been known to be
excellent for the four-jet rate in $e^+e^-$
annihilation~\cite{NLOZ4Jets}.  A similar approximation was used
recently for an investigation of \Wjjj-jet production~\cite{EMZ},
which, however, also omitted many partonic subprocesses.  Our study
retains all subprocesses.  In addition, we keep all
subleading-color terms in the real-emission contributions.  In the
finite virtual terms of each subprocess we drop certain
subleading-color contributions.  ``Finite'' refers to the $\e^0$ term
in the Laurent expansion of the infrared-divergent one-loop amplitudes
in $\e = (4-D)/2$, after extracting a multiplicative factor of
$c_\Gamma(\e) \equiv
\Gamma(1+\e)\Gamma^2(1-\e)/\Gamma(1-2\e)/(4\pi)^{2-\e}$.
``Subleading-color'' refers to the part of the ratio of the virtual
terms to tree cross section that is suppressed by at least one power
of either $1/N_c^2$ or $n_f/N_c$ (virtual quark loops).  We multiply
the surviving, leading-color terms in this ratio back by the tree
cross section, with its full color dependence.

For this approximation, we need only the color-ordered
(primitive) amplitudes in which the $W$ boson is adjacent to the two
external quarks forming the quark line to which it
attaches. Representative Feynman diagrams for these primitive
amplitudes are shown in \fig{lcdiagramsFigure}.  Other primitive
amplitudes have external gluons (or a gluon splitting to a $\bar Q Q$ pair)
attached between the $W$ boson and the
two above-mentioned external quarks; they only contribute~\cite{qqggg}
to the subleading-color terms that we drop.  As discussed below, we
have confirmed that for \Wjx-jet production this leading-color
approximation is valid to within three percent, so we expect
corrections to the \Wjjj-jet cross-sections from subleading-color
terms also to be small. 

In addition to the virtual corrections to the cross section
provided by \BlackHat, the NLO result also requires the real-emission
corrections to the LO process.  The latter arise from
tree-level amplitudes with one additional parton, either an
additional gluon, or a quark--antiquark pair replacing a gluon.
Infrared singularities develop when the extra parton momentum
is integrated over unresolved phase-space regions.
They cancel against singular terms in the virtual
corrections, and against counterterms associated with the evolution of
parton distributions.  We use the
program~\AMEGIC~\cite{Amegic} to implement these cancellations
via the Catani-Seymour dipole subtraction method~\cite{CS}.  The \SHERPA{}
framework~\cite{Sherpa} incorporates \AMEGIC, making it easy to
analyze the results and construct a wide variety of distributions.
For other automated implementations of the dipole subtraction
method, see refs.~\cite{AutomatedSubtractionOther}.

The CDF analysis~\cite{WCDF} employs the {\sc JETCLU} cone
algorithm~\cite{JETCLU} with a cone radius $R = \sqrt{(\Delta \phi)^2
+ (\Delta \eta)^2} = 0.4$.  However, this algorithm is not generally
infrared safe at NLO, so we instead use the seedless cone algorithm
{\sc SISCone}~\cite{SISCONE}.  In general, at the partonic level we
expect similar results from any infrared-safe cone algorithm.  For
\Wjx{} jets we have confirmed that distributions using {\sc SISCone}
are within a few percent of those obtained with the midpoint cone
algorithm~\cite{Midpoint}.

Both electron and positron final states are counted, and the following
cuts are imposed: $E_T^{e} > 20$ GeV, $|\eta^e| < 1.1$, $\ETsl > 30$
GeV, $M_T^W > 20$ GeV, and $E_T^{\rm jet} > 20$ GeV.  Here $E_T$ is
the transverse energy, $\ETsl$ is the missing transverse energy,
$M_T^W$ the transverse mass of the $e \nu$ pair and $\eta$ the
pseudorapidity.  Jets are ordered by $E_T$, and are required to have
$|\eta| < 2$.  Total cross sections are quoted with a tighter jet cut,
$E_T^{\rm jet} > 25$ GeV.  CDF also imposes a minimum $\Delta R$ between the
charged decay lepton and any jet; the effect of this cut, however, is
removed by the acceptance corrections.

\begin{table}
\vskip .4 cm
\begin{tabular}{|c||c|c|c|}
\hline
number of jets  & CDF &  LC NLO & NLO  \\
\hline
1  & $\; 53.5 \pm 5.6 \;$ & $\; 58.3^{+4.6}_{-4.6} \;$ & 
            $\;  57.8^{+4.4}_{-4.0} \;$ \\
\hline
2  & $6.8 \pm 1.1$  & $ 7.81^{+0.54}_{-0.91}$ &
            $7.62^{+0.62}_{-0.86} $  \\
\hline
3 &  $0.84\pm 0.24$  & $\;0.908^{+0.044}_{-0.142} \;$ & ---  \\
\hline 
\end{tabular} 
\caption{Total cross sections in pb for \Wjn{} jets with $E_T^{n\rm
th\hbox{-}jet} > 25$ GeV as measured by CDF~\cite{WCDF}. The results
are compared to NLO QCD. For \Wj{} and \Wjj{} jets, the
difference between the leading-color approximation and the complete
NLO result is under three percent.  For \Wjjj{} jets only the LC NLO
result is currently available, but we expect a similarly small
deviation for the full NLO result.  Experimental statistical,
systematic and luminosity uncertainties have been combined for the CDF
results.
\label{CDFCrossSectionTable} }
\end{table}

\begin{figure*}[t]
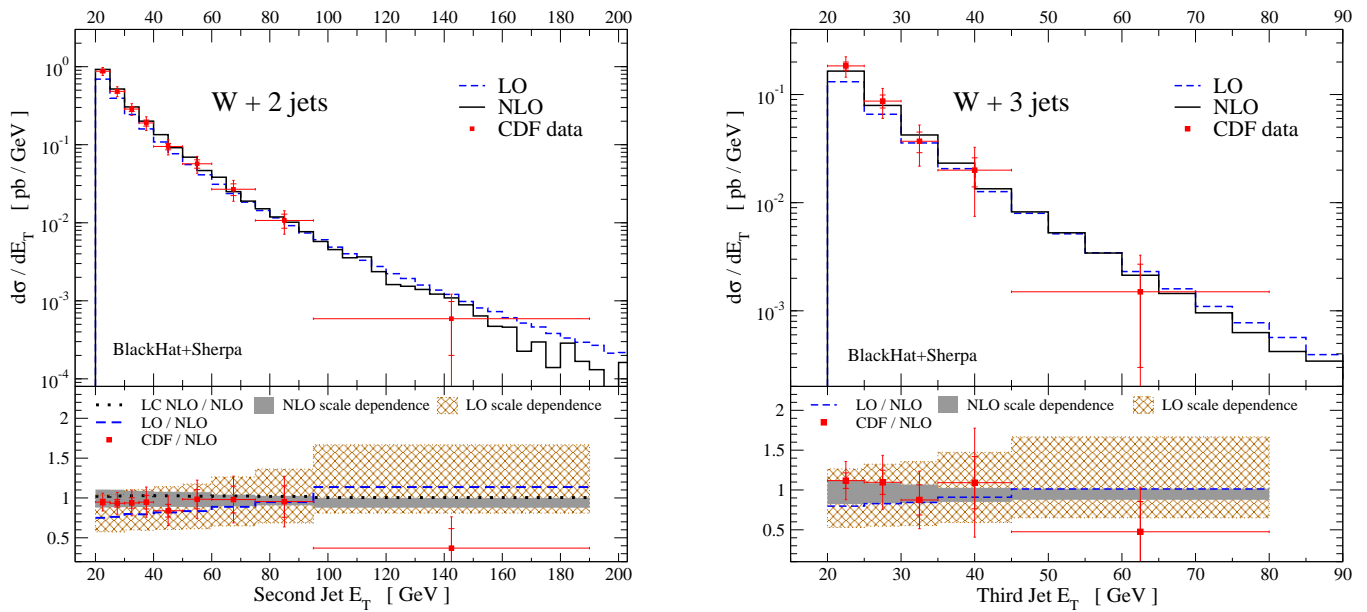

\includegraphics[clip,scale=0.4]{W2j_Et_sub_lead_jet.eps}\ \ 
\hskip 1 cm
\includegraphics[clip,scale=0.4]{W3j_Et_sub_sub_lead_jet.eps}

\caption{ The measured cross section $d \sigma(W\rightarrow e \nu +
\ge n$-jets)$/dE_T^{n\rm th\hbox{-}jet}$ compared to NLO predictions
for $n=2,3$.  In the upper panels the NLO distribution is the solid
(black) histogram, and CDF data points are the (red) points, whose
inner and outer error bars denote the statistical and total
uncertainties on the measurements.  The LO predictions are shown as
dashed (blue) lines.  The lower panels show the distribution
normalized to an NLO prediction, the full one for $n=2$ and the
leading-color one for $n=3$, in the experimental bins (that is,
averaging over several bins in the upper panel).  The scale
uncertainty bands are shaded (gray) for NLO and cross-hatched (brown)
for LO.  In the $n=2$ case, the dotted (black) line shows the ratio of
the leading-color approximation to the full-color calculation. }
\label{W23Figure}
\end{figure*}

CDF compared~\cite{WCDF} their measured \Wjn-jet cross sections to LO
(matched to partons showers~\cite{MLMSMPR}) and the then-available NLO
theoretical predictions.  The LO calculations differ substantially
from the data, especially at lower $E_T$, and have large
scale-dependence bands.  In contrast, the NLO calculations for $n\le
2$ jets (using the MCFM code~\cite{MCFM}, with the $V+4$-parton
one-loop matrix elements of ref.~\cite{Zqqgg}) show much better
agreement, and narrow scale-dependence bands. See ref.~\cite{WCDF} for
details.

Our aim in this Letter is to extend this comparison to $n = 3$ jets.
We apply the same lepton and jet cuts as CDF, replacing the $\ETsl$
cut by one on the neutrino $E_T$, and ignoring the lepton--jet $\Delta
R$ cut removed by acceptance.  We approximate the
Cabibbo-Kobayashi-Maskawa matrix by the unit matrix, express the $W$
coupling to fermions using the Standard Model parameters $\alpha_{\rm
QED} =1/128.802$ and $\sin^2\theta_W=0.230$, and use $m_W = 80.419$
GeV and $\Gamma_W=2.06$ GeV.  We use the CTEQ6M~\cite{CTEQ6M} parton
distribution functions (PDFs) and an event-by-event common
renormalization and factorization scale, $\mu = \sqrt{m_W^2 + p_T^2(W)
}$.  To estimate the scale dependence we choose five values in the
range $({1\over 2}, 2)\times \mu$.  The numerical
integration errors are on the order of a half percent.
 We do not include PDF
uncertainties.  For \Wjx-jet production these uncertainties have been
estimated in ref.~\cite{WCDF}.  In general they are smaller than the
scale uncertainties at low $E_T$ but larger at high $E_T$. The LO
calculation uses the CTEQ6L1 PDF set.  For $n=1,2$ jets, NLO total
cross sections agree with those from MCFM~\cite{MCFM}, for various
cuts.  As our calculation is a parton-level one, we do not apply
corrections due to non-perturbative effects such as induced by the
underlying event or hadronization. Such corrections are expected to be
under ten percent~\cite{WCDF}.

In \tab{CDFCrossSectionTable}, we collect the results for the total
cross section, comparing CDF data to the NLO theoretical predictions
computed using \BlackHat{} and \SHERPA{}.  The columns labeled ``LC
NLO'' and ``NLO'' show respectively the results for our leading-color
approximation to NLO, and for the full NLO calculation.  The
leading-color NLO and full NLO cross-sections for \Wj- and \Wjj-jet
production agree to within three percent. We thus expect only a small
change in the results for \Wjjj-jet production once the missing
subleading-color contributions are incorporated.

We have also compared the $E_T$ distribution of the $n^{\rm th}$ jet
in CDF data to the NLO predictions for \Wjjjx-jet production.  For
\Wjjx-jets these comparisons are shown in fig.~\ref{W23Figure},
including scale-dependence bands obtained as described above.  For
reference, we also show the LO distributions and corresponding
scale-dependence band.  (The calculations matching to parton
showers~\cite{MLMSMPR} used in ref.~\cite{WCDF} make different choices
for the scale variation and are not directly comparable to the
parton-level predictions shown here.)  The NLO predictions match the
data very well, and uniformly in all but the highest $E_T$ bin.  The
central values of the LO predictions, in contrast, have different
shapes from the data.  The scale dependence of the NLO predictions is
substantially smaller than that of the LO ones.  In the \Wjj-jet case,
we also show the ratio of the leading-color approximation to the
full-color result within the NLO calculation: the two results differ
by less than three percent over the entire transverse energy range,
considerably smaller than the scale dependence (and experimental
uncertainties).

\begin{figure}[t]
\includegraphics[clip,scale=0.4]{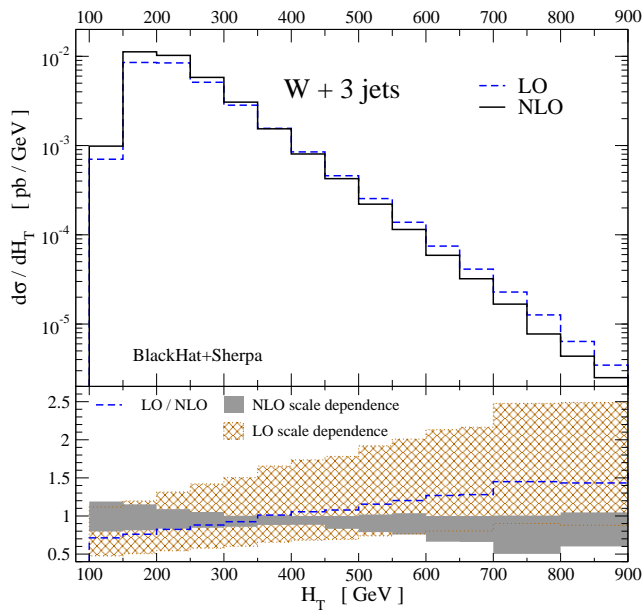}
\caption{The theoretical prediction for the $H_T$ distribution
in \Wjjj-jet production. The curves and bands
are labeled as in \fig{W23Figure}. }
\label{SecondDistFigure}
\end{figure}

In \fig{SecondDistFigure}, we show the distribution for the total
transverse energy $H_T$, given by the scalar sum of the jet and lepton
transverse energies, $H_T = \sum_j E_{T,j}^{\rm jet} + E_T^e +
\ETsl$.  We show the NLO and LO predictions, along with their
scale-uncertainty bands. As in the $E_T$ distributions, the NLO band
is much narrower; and the shape of the distribution is altered at NLO
from the LO prediction.

In summary, we have presented the first phenomenologically useful NLO
study of \Wjjj-jet production, and compared the total cross section
and the jet $E_T$ distribution to Tevatron data~\cite{WCDF}.  The
results demonstrate the utility of the on-shell method and its
numerical implementation in the \BlackHat{} code for NLO computations
of phenomenologically-important processes at the LHC.


\vskip .3 cm 
We thank Jay Hauser, Warren Mori, Sasha Pronko and Rainer
Wallny for helpful discussions.  This research was supported by the US
Department of Energy under contracts DE--FG03--91ER40662,
DE--AC02--76SF00515 and DE--FC02--94ER40818.  DAK's research is
supported by the Agence Nationale de la Recherce of France under grant
ANR--05--BLAN--0073--01, and by the European Research Council under
Advanced Investigator Grant ERC--AdG--228301.  This research used
resources of Academic Technology Services at UCLA and of the National
Energy Research Scientific Computing Center, which is supported by the
Office of Science of the U.S. Department of Energy under Contract
No. DE-AC02-05CH11231.

\end{document}